\documentclass[12pt,preprint]{aastex}

\usepackage{url}
\usepackage{color}

\shorttitle{Error corrected Gaussian Mixture}

\shortauthors{Hao et al.}

\begin{document}


\title{Precision Measurements of the Cluster Red Sequence using an Error Corrected Gaussian Mixture Model}

\author{Jiangang Hao\altaffilmark{1,2}, Benjamin P. Koester\altaffilmark{3},
Timothy A. Mckay \altaffilmark{2,4}, Eli S. Rykoff\altaffilmark{5},
Eduardo Rozo\altaffilmark{6}, August Evrard\altaffilmark{2,4}, James
Annis\altaffilmark{1}, Matthew Becker\altaffilmark{7}, Michael
Busha\altaffilmark{8}, David Gerdes\altaffilmark{2}, David E.
Johnston\altaffilmark{9}, Erin Sheldon\altaffilmark{10}, Risa H.
Wechsler\altaffilmark{8}}

\altaffiltext{1}{Center for Particle Astrophysics, Fermi National Accelerator Laboratory, Batavia, IL 60510}
\altaffiltext{2}{Department of Physics, University of Michigan,
Ann Arbor, MI 48109}
\altaffiltext{3}{Department of Astronomy and
Astrophysics, The University of Chicago, Chicago, IL60637}
\altaffiltext{4}{Department of Astronomy, University of Michigan,
Ann Arbor, MI48109}
\altaffiltext{5}{TABASGO Fellow, Physics Department, University of
California at Santa Barbara, 2233B Broida Hall, Santa Barbara, CA
93106}
\altaffiltext{6}{Center for Cosmology and Astro-Particle
Physics (CCAPP), The Ohio State University, Columbus, OH 43210}
\altaffiltext{7}{Department of Physics, The University of Chicago,
Chicago, IL 60637}
\altaffiltext{8}{Kavli Institute for Particle Astrophysics \&
Cosmology, Physics Department, and Stanford Linear Accelerator
Center, Stanford University, Stanford, CA 94305}
\altaffiltext{9}{Department Physics \& Astronomy, Northwestern University, Evanston, IL 60208}
\altaffiltext{10}{Brookhaven National Laboratory, Upton, New York 11973, USA}


\begin{abstract}

The red sequence is an important feature of galaxy clusters and
plays a crucial role in optical cluster detection. Measurement of
the slope and scatter of the red sequence are affected both by
selection of red sequence galaxies and measurement errors. In this
paper, we describe a new error corrected Gaussian Mixture Model for
red sequence galaxy identification. Using this technique, we can
remove the effects of measurement error and extract unbiased
information about the intrinsic properties of the red sequence. We
use this method to select red sequence galaxies in each of the
13,823 clusters in the maxBCG catalog, and measure the red sequence
ridgeline location and scatter of each. These measurements provide
precise constraints on the variation of the average red galaxy
populations in the observed frame with redshift. We find that the
scatter of the red sequence ridgeline increase mildly with redshift,
and that the slope decreases with redshift. We also observe that the
slope does not strongly depend on cluster richness. Using similar
methods, we show that this behavior is mirrored in a spectroscopic
sample of field galaxies, further emphasizing that ridgeline
properties are independent of environment. These precise
measurements serve as an important observational check on
simulations and mock galaxy catalogs. The observed trends in the
slope and scatter of the red sequence ridgeline with redshift are
clues to possible intrinsic evolution of the cluster red-sequence
itself. Most importantly, the methods presented in this work lay the
groundwork for further improvements in optically-based cluster
cosmology. The codes for ECGMM can be accessed from: \color{blue}\url{https://sites.google.com/site/jiangangecgmm/}

\end{abstract}




\section{Introduction}
Galaxy clusters are the largest gravitationally bound systems in our
Universe, whose masses, abundance and spatial distribution reflect
the growth of structure, composition, and expansion history of the
Universe~\citep{evrard89,oukbir92}. The feasibility of constraining
cosmological parameters using galaxy clusters has been demonstrated
by many authors~\citep{mohr04self,hu04self,lima04,lima05} and
realistic constraints on cosmological parameters from optically
selected galaxy clusters have been implemented recently by
~\citet{gladders07} and~\citet{rozo09} on the RCS cluster
catalog~\citep{gladders05rcs} and maxBCG
catalog~\citep{koester07cat,koester07alg} respectively.

The predominantly red, bright, passively evolving red sequence, or
``E/S0 ridgeline''~\citep{visvanathan77,annis99} found in the cores
of clusters of varied richness up to at least $z \sim 1.4$
~\citep{bower92,smail98,vandokkum98, barrientos99,blakeslee03,
mullis05,eisenhardt05,delucia07} provide an efficient means for
cluster detection, and have been an integral part of modern cluster
cosmology. The red sequence itself is ubiquitous in the galaxy
population~\citep[e.g.]{renzini06}, and in clusters red sequence
galaxies dominate the bright end of the cluster luminosity function
~\citep{sandage85,barger98}. They are extremely tightly clustered in
color space, containing old populations of stars whose observed
color varies smoothly with redshift ~\citep[e.g.][]{gladders00}. The
pervasiveness of this phenomenon in clusters enables efficient
optical cluster detection while a fortuitous color-redshift relation
yields accurate photometric redshifts~\citep{gladders00,
koester07alg}. Simple counting of photometrically identified cluster
red sequence galaxies ~\citep[e.g]{koester07cat} has also been shown
to be an effective proxy for cluster
mass~\citep{becker07,sheldon07,johnston07}, with more sophisticated
applications yielding improvements in richness as a cluster mass
proxy~\citep{rozo08}. In the era of precision measurements, the
extent to which the red sequence can be exploited for cluster
cosmology depends on how accurately its characteristics can be
measured at a given redshift.

In addition to its relevance to cluster cosmology, the red sequence
plays an important role in constraining the complex physical
processes that drive galaxy formation and evolution. At the field
scale this includes measurements of the red galaxy luminosity
function~\citep{wake06,faber07}, the clustering of red galaxies in
various environments~\citep{zehavi05,coil08}, and color-magnitude
relations of spectroscopically and morphologically identified early-type galaxies~\citep{cool06,mei06,stanford98}. The high density environments of clusters of galaxies are dominated by red sequence galaxies; the red
sequence portion of the galaxy populations in the cores of rich
clusters to at least $z \sim 1$ form the basis for various
monolithic collapse scenarios
~\citep[e.g.][]{bower92,blakeslee03,mei09}.~\citet{faber07}
summarize some of these results to fill out a picture of galaxy
formation that includes a mechanism for the formation of the red
sequence.

In color-magnitude space, the red sequence is typically
characterized by its slope, zero point, and scatter. Many theoretical modeling have been proposed to explain the red sequence~\citep[e.g.][]{arimoto87,kauffmann98}. Various models
posit that in the rest frame, the scatter in the red sequence is
driven primarily by age effects, its slope is a manifestation of the
mass-metallicity relation, and the zero point is set by combination
of age and mass-metallicity differences
~\citep[e.g.][]{bernardi05,delucia07,faber07}.

Studies of the cluster red sequence have been accomplished by simply
measuring the photometric color-magnitude relation
~\citep[e.g.][]{lcyee04,delucia07,stott09}, supplemented with HST
morphological information~\citep[e.g][]{gladders98} and spectroscopy
at higher redshift \citep{mei06,stott09}. Extra morphological and spectroscopic data allow
precise separation of E and S0-types from the rest of the galaxy
population, as well as refined identification of cluster members
~\citep{blakeslee03,mei09}. The situation also benefits
significantly from precise color measurements afforded by deep,
CCD-based imaging~\citep[e.g][]{vandokkum98}. In the literature, the
red sequence has been measured with various levels of scrutiny in
dozens of individual clusters.

In the past several years, researchers have turned to the
considerable resources of the the Sloan Digital Sky Survey (SDSS)
and similar wide field surveys to probe red sequence and elliptical
galaxies in various
environments~\citep{hogg04,bernardi05,bernardi06,cool06}. These
studies include both spectroscopically and morphologically
identified red galaxies at $z \sim 0.1$ that aim to constrain galaxy
evolution scenarios for the cosmologically-relevant luminous red
galaxy (LRG) samples that extend to $z \sim 0.6$
~\citep[e.g.][]{cool06}.

With the maxBCG cluster catalog, we are positioned to use the SDSS
to make one of the most statistically robust photometric measurements
of the cluster red sequence, using nearly 14,000 clusters between
$0.1 \le z \le 0.3$. In this paper we focus on the slope and scatter
of the red sequence. We clearly show the systematic effects photometric errors
have on the measurement of the underlying slope and scatter of the
red sequence, and introduce a method for properly handling these
effects. This method, based on an error-corrected Gaussian Mixture
Model (ECGMM), reliably recovers the properties of the ridgeline by
taking measurement errors into account. After presenting the method,
we describe its application to measurement of maxBCG clusters. Of
particular relevance to cluster cosmology are the observed mean,
scatter, and slope of the E/S0 ridgeline for all maxBCG clusters.
These results are presented, along with a discussion of observed
trends with redshift.


\section{Methods}

\subsection{Intrinsic properties of red sequence ridgeline}

The existence of the red sequence ridgeline is evidence that cluster
galaxies are clustered in color space in addition to real space. The
emission from early-type galaxies is dominated by old stellar
populations, which gives rise to these remarkably similar galaxy
colors. In addition, there is a close mapping between galaxy color
and redshift for these galaxies as a result of the restframe
4000~\AA~ break in their spectra. For the SDSS filter sets, the
4000~\AA~ break is within the $g$ band as long as the redshift is
below 0.35. Therefore, the most informative ridgeline color for the
maxBCG catalog is $g-r$.

In the projected vicinity of a detected cluster, there are both
cluster member galaxies and field galaxies. Red sequence galaxies
form a part of the member population, whose colors are clustered
tightly and can be approximated with a Gaussian distribution with
narrow width. On the other hand, the field galaxies' and blue member
galaxies' colors are not tightly clustered and can be approximated
by a Gaussian distribution with a broader width\footnote{There are
complicated situations where the distribution in color space is not
simply unimodal or bimodal, for example when two clusters are seen
in projection. For maxBCG catalog, it covers about 7,500 square degrees with about 14,000 clusters. This leads to about 2 clusters per square degree. Each cluster is about the size of a few arcminutes across, so the chance of two or more overlapped clusters is low. Therefore, a
unimodal or bimodal distribution in color space is a good
approximation.}. The problem of separating the ridgeline from the
field can be specified as following: What are the two Gaussian
components (one for the ridgeline and one for the field) that
represent the color distribution in the vicinity of a galaxy
cluster? If this double Gaussian is an adequate model for describing
this color distribution, the one dimensional Gausian Mixture Model
(GMM) is well suited to the problem.

In the traditional applications of GMM~\citep{gmmbook}, measurement
errors are not considered. In our case, there are non-negligible
measurement errors associated with the galaxy colors. We are
interested in measuring the intrinsic color scatter of cluster
members, absent contamination by the increasing measurement errors
of faint galaxies. Without accounting for the increasing photometric
errors, we expect that the color scatter will increase as the
measurement errors become larger. While the intrinsic color scatter
may increase as redshift increases (because the 4000 \AA~ line break
is shifting toward $r$ band and making the $g-r$ color less
discriminative.), measurement errors may make us overestimate the
increase in intrinsic scatter with redshift. To avoid this problem,
we include measurement error into our likelihood function to remove
the contamination. We will refer to this as the error-corrected
Gaussian Mixture Model and derive the corresponding Expectation
Maximization (EM) recursive relation in the following section.

It is clear that we can always improve the fit by adding more Gaussian
components, although this is clearly not good in the sense of
parsimony. So, we need to somehow decide on the number of Gaussian
components by trading off quality of fit against the number of
introduced free parameters. To do so, we use the Bayesian
Information Criterion (BIC)~\citep{schwarz78,connolly00} to
determine how many mixtures we should use. The BIC is defined as:

\begin{equation}
BIC=-2\log\mathcal{L}_{max}+k\log(M)
\end{equation}

\noindent Where k is the number of free parameters. For mixture models with different number of components, we
compare their corresponding BIC and select the model with the
smallest BIC.

\subsection{Error-corrected Gaussian Mixture Model}

In what follows, we describe how to fit a multi-component Gaussian
mixture model to a one dimensional distribution of data with both
intrinsic scatter and Gaussian measurement errors. Our method is an
extension of the traditional expectation maximization method for
GMM~\citep{em77}.

We assume the data are to be modeled by a mixture of $N$ Gaussians
fit to the distribution of $M$ data points. The subscript $i$ cycles
through $N$ and $j$ cycles through $M$, and we use $\mu_i$,
$\sigma_i$ and $w_i$ to denote the location, width and weight of
each Gaussian component. $y_j$ and $\delta_j$ denote the data points
and their measurement errors which are assumed to be Gaussian. For
brevity, we denote the parameters ($\mu_i$, $\sigma_i$ and $w_i$)
collectively by $\theta$. Then the likelihood of the parameters
given the data and measurement errors is:

\begin{equation}\label{totallk}
\mathcal{L}(\theta|y)=\prod_{j=1}^M
\bigg\{\sum_{i=1}^N\frac{w_i}{\sqrt{2\pi(\sigma_i^2+\delta_j^2)}}
\exp\bigg[-\frac{(y_j-\mu_i)^2}{2(\sigma_i^2+\delta_j^2)}\bigg]\bigg\}
\end{equation}

The optimal parameters $\theta$ could be estimated by maximizing the
above likelihood function. The Expectation Maximization algorithm
provides an efficient way to get the maximum likelihood estimators
in such a setting. To utilize this, we need to introduce a hidden
variable, $z_j$, which tells which Gaussian component the data point
$y_j$ is sampled from. In our case, different from the standard EM
prescription, we have non-negligible measurement errors present.
After some algebra, we arrive at a set of recursive relations that
lead to the maximum of the likelihood (see appendix for details).

\subsection{Bootstrapping to increase the robustness of ECGMM}

Though the ECGMM is generally stable for estimating the parameters
of the Gaussian Components, it can fail occasionally due to very
inappropriate choice of initial parameters or some very big
measurement errors of certain galaxies. To make our measurement more
robust, we introduced a bootstrap-like scheme. Suppose we have M
data points. We randomly pick one of the data points and record it.
We then repeat this process M times and get M recorded data points.
These M points form one resampling set of the original data set.
Now, we apply the ECGMM to this new data sample and measure the
corresponding parameters. After this, we start a second round,
getting another resampling set with M data points in it and measure
the parameters using ECGMM again. We repeat this process X times and
have X estimates of each parameters. We throw away those outlier
estimates (estimates beyond the upper and lower inner
fences\footnote{In statistics, lower inner fence is defined by
$Q_1-1.5IQR$ and higher inner fence is $Q_3+1.5IQR$, where $Q_1$ and
$Q_3$ are the first and third quartiles respectively. The IQR is the
the interquartile range, defined as $Q_3-Q_1$.}) for each parameter
and use the mean of good estimates as the value of each parameter.
Using this scheme, our resulting parameter estimates are much more
robust, at a cost of a tolerable increase in computation time. In
this application, we took X to be 50.

\subsection{Monte Carlo test of the ECGMM for our application}

Before we delve into real data, we first conduct Monte Carlo tests
to see whether the ECGMM approach can reliably identify the cluster
and background Gaussian components. These tests are used to
determine whether this method can reliably recover the true
parameters input in the simulation, and to see whether the extracted
parameters are generally unbiased with respect to measurement
errors.

\begin{figure*}
\epsscale{0.9} \plotone{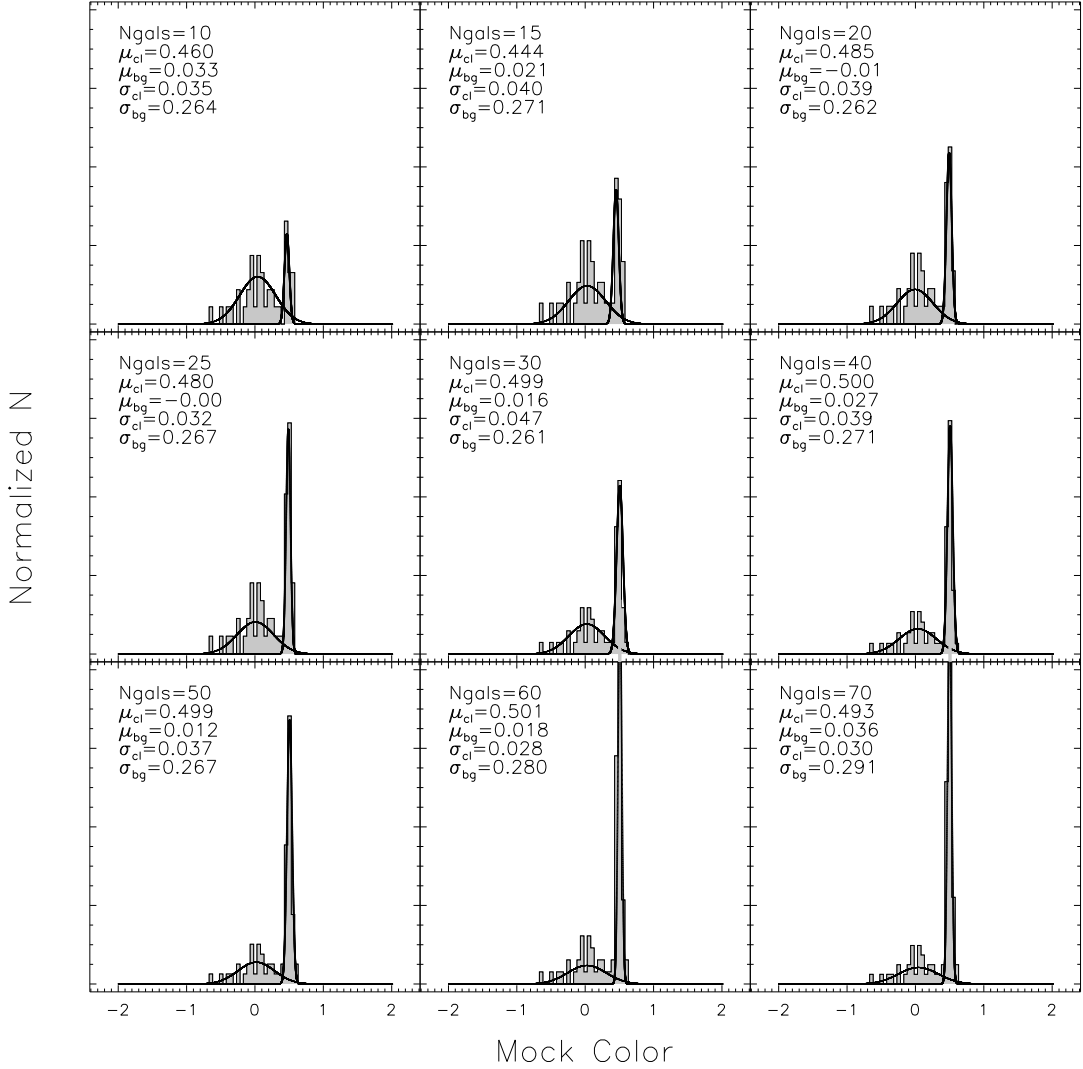} \caption{Basic Monte Carlo tests
of the ECGMM method for fitting mock galaxy $g-r$ color
distributions (see text). $\mu $ and $\sigma$ denote the locations
and widths of the corresponding Gaussian components, for clusters of
increasing richness. The true $\mu$ are 0 and 0.5 for BG and CL sets
respectively. The true $\sigma$ are 0.3 and 0.04 for BG and CL sets
respectively. } \label{fig:mccal}
\end{figure*}

For this purpose, we generate two Gaussian random data sets, one
representing cluster member colors, denoted CL, and the other
representing the field galaxies/blue galaxies' colors, denoted as
BG. The CL set is generated from $\sim N(0.5,0.04^2)$ and BG set is
generated from $\sim N(0,0.3^2)$. To represent clusters with
different richness, we allow the normalization (also denoted as
$N_{gals}$ in the plots) of CL data set to vary as 10, 15, 20, 25,
30, 40, 50, 60 and 70 while keep the normalization of BG set as 30. All the parameters used to generate the mock data are chosen to make the simulation as close to the real data as possible. Then we combine CL and BG to create a mock data
set that mimics the colors of both cluster members and background
galaxies in a field. It is worth noting that these mock colors are
error free so far. Next, we will add some noise to them to mimic the
measurements errors. To do this, we first generate random numbers
from a uniform distribution in the range of [0, 0.1], which play the
role of $\delta_j$ in Eq.\ref{totallk}. Then, we generate from $N(0,
\delta_j^2)$ and add them to the noise free data set to produce a
noise added mock color data set. In Fig.\ref{fig:mccal}, we plots
the results from the ECGMM fitting. The results show that for
clusters with $N_{gals} \ge 10$, the method can recover the
locations ($\mu$) and widths ($\sigma$) of the Gaussian components very well.

\begin{figure*}[tb]
\epsscale{0.6} \plotone{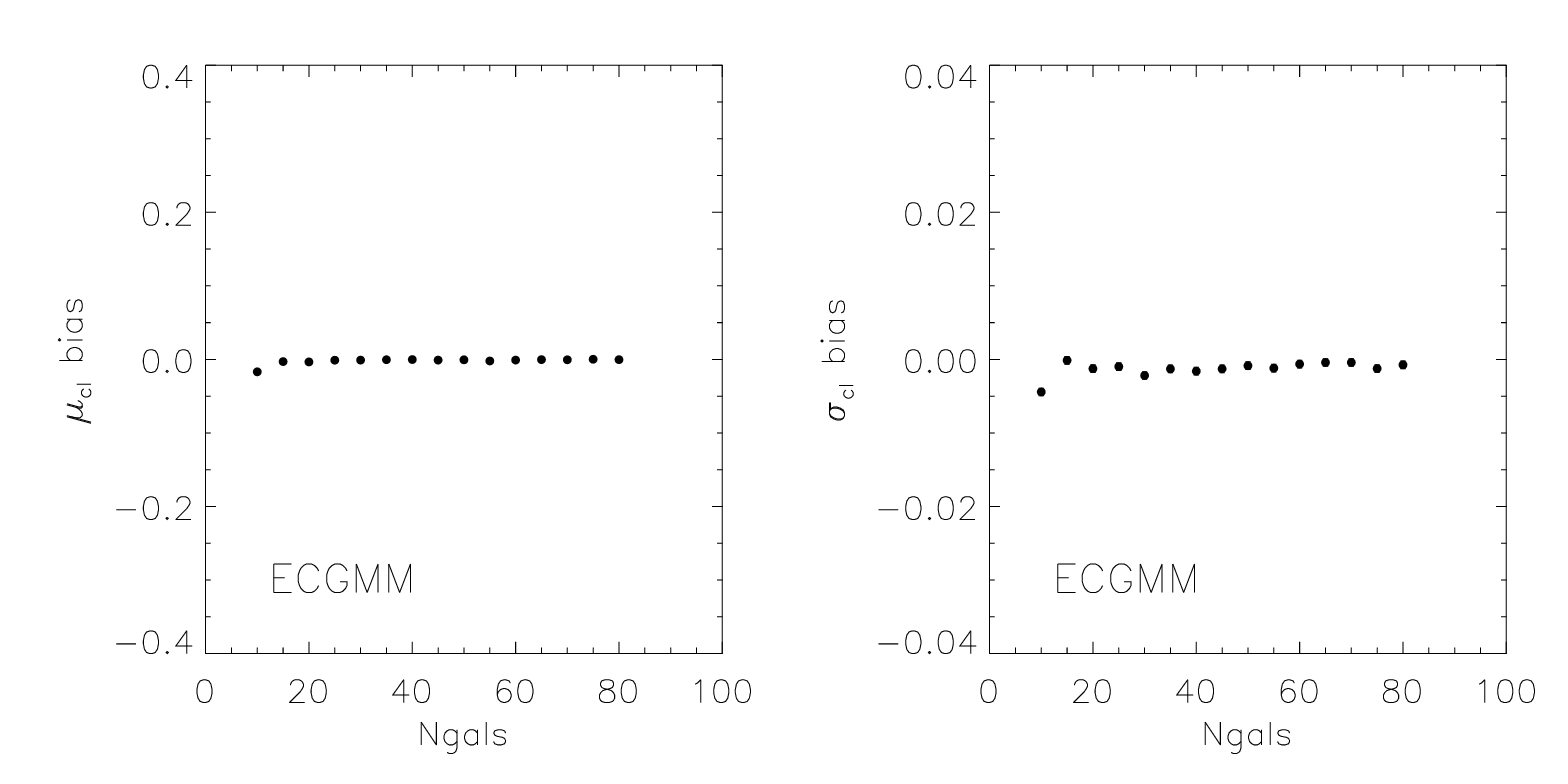}
\plotone{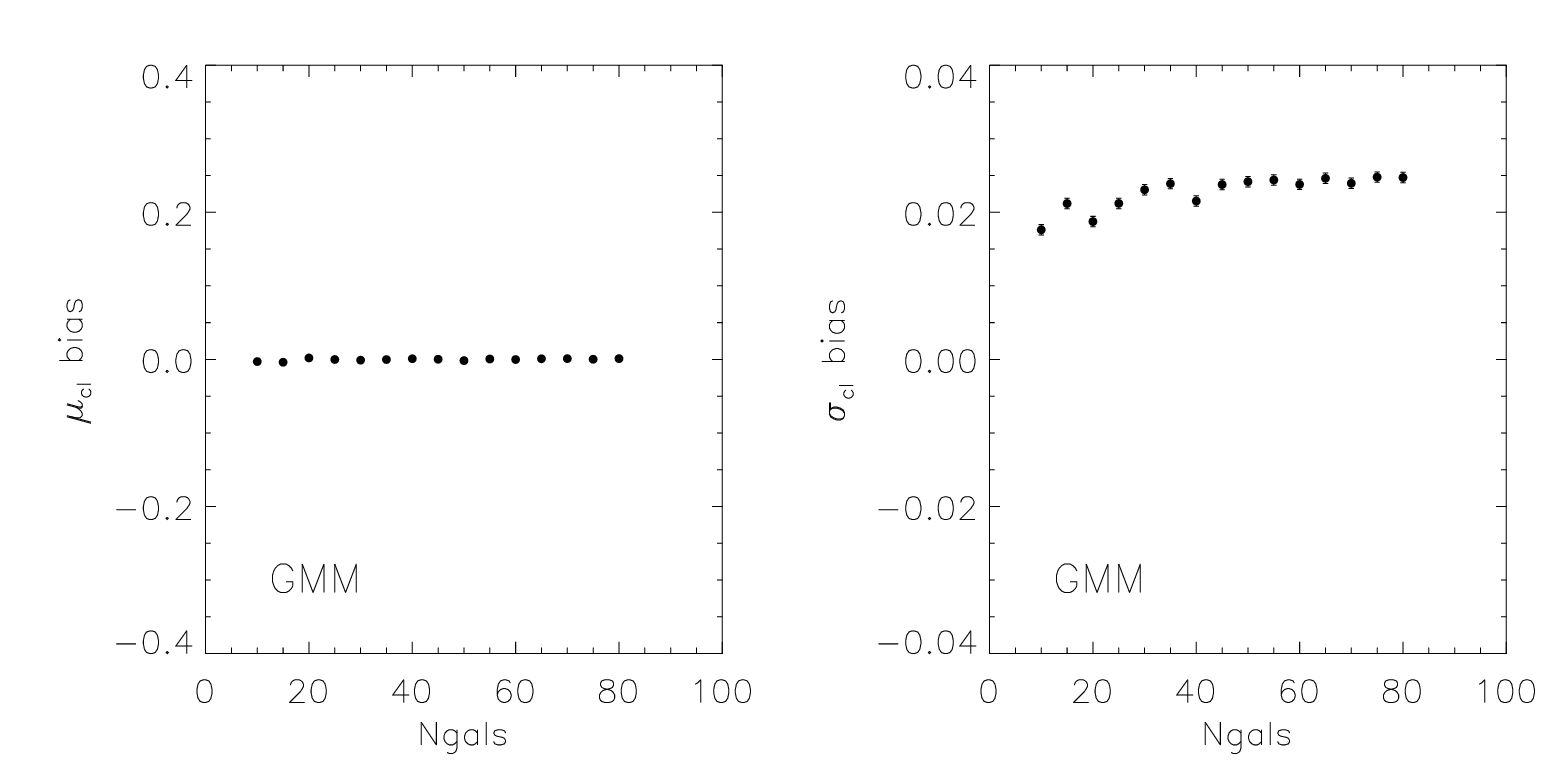} \caption{Monte Carlo test of the
bias, ($ E(\hat{\theta})-\theta$), of the estimators for the
location and width using GMM (bottom two panels) and ECGMM (top two
panels) as a function of richness for the cluster component of the
mock clusters (see text). The scales of the plots are chosen to be
close to the size of the true parameters in the plot to illustrate
the fraction precision.} \label{fig:mcbias}
\end{figure*}



Next we test for possible bias in the estimators. For each cluster
richness $N_{gals}$, we re-generate the data as well as errors 200
times and then apply our methods to each to obtain estimates for the
parameters. In each case, we calculate the bias of parameters
$\theta$ (the $\sigma$ and $\mu$ in our case) defined as $
E(\hat{\theta})-\theta$. In Fig.\ref{fig:mcbias}, we plot the
results from both GMM and ECGMM for comparison. Clearly, the
introduction of error correction(as shown in the bottom two panels)
is essential for removal of the bias of the width resulting from
measurement error(as shown in the top two panels).


\section{Data}
\subsection{SDSS}
Three main resources are included in this work: the SDSS galaxy
catalog, the maxBCG catalog, and a value-added SDSS spectroscopic
catalog.

The maxBCG cluster sample and the galaxy catalogs used to remeasure
cluster richness in this paper are derived from the SDSS~\citep{adelman06}. The
maxBCG cluster sample covers a sky area of about 7500 square degrees. The camera design~\citep{gunn06} and drift-scan imaging strategy of the SDSS enable acquisition of nearly
simultaneous observations in the $u,g,r,i,z$ filter system
~\citep{fukugita96}. Calibration~\citep{hogg01,smith02,tucker06},
astrometric~\citep{pier03}, and photometric~\citep{lupton01}
pipelines reduce the data into object catalogs containing a host of
measured parameters for each object. Galaxies are selected from SDSS
object catalogs as described in~\citep{sheldon07}. In this work we
use $\tt{CMODEL\_COUNTS}$ as our total magnitudes, and
$\tt{MODEL\_COUNTS}$ when computing colors. Bright stars, survey
edges and regions of poor seeing are masked as previously described
~\citep{koester07cat,sheldon07}.


The spectroscopic galaxy catalog is comprised of galaxies from the
DR6 of SDSS Value Added Galaxy Catalog. A detailed description about
this catalog can be found in~\citet[VAGC]{blantonvagc}.

\subsection{Cluster Sample}

We obtain sky locations, redshift estimates, and initial richness
values from the maxBCG cluster catalog. Details of the selection
algorithm and catalog properties are published
elsewhere~\citep{koester07alg,koester07cat}. In brief, maxBCG
selection relies on the observation that the galaxy population of
rich clusters is dominated by luminous, red galaxies clustered
tightly in color (the E/S0 ridgeline). Since these galaxies have
old, passively evolving stellar populations, their $g-r$ color
closely reflects their redshift.  The brightest such red galaxy,
typically located at the peak of the galaxy density, defines the
cluster center.

The maxBCG catalog is approximately volume limited in the redshift
range $0.1 \le z \le 0.3$, with very accurate photometric redshifts
($\delta{}z \sim 0.01$). Studies of the maxBCG algorithm applied to
mock SDSS catalogs indicate that the completeness and purity are
very high, above $90\%$~\citep{koester07cat}. The maxBCG catalog has
been used to investigate the scaling of galaxy velocity dispersion
with cluster richness~\citep{becker07} and to derive constraints on
the power spectrum normalization, $\sigma_8$, from cluster number
counts~\citep{rozo09}.

\section{Measuring the ridgeline location and width of maxBCG clusters}

We apply the above prescriptions of ECGMM to the maxBCG cluster
catalog and the galaxy catalog~\citep{koester07cat}, measuring the
red sequence $g-r$ ridgeline. The procedures are as follows: for
each cluster in maxBCG catalog, we choose a scaled aperture
$R_{200}^{lens}$ to ensure we are considering equivalent regions of
clusters of varied masses and therefore varied richness.
$R_{200}^{lens}$ is the critical radius, interior to which the mean
mass density of the cluster is 200 times of the critical energy
density. Based on the weak lensing
analysis~\citep{johnston07,hansen07}, the scaling relation between
$R_{200}^{lens}$ and the original maxBCG richness $N_{200}$ is given
by $R_{200}^{lens} = 0.182(N_{200})^{0.42}$, which ranges from 0.47 Mpc to 1.68 Mpc.

\begin{figure}
\epsscale{.9}
\plotone{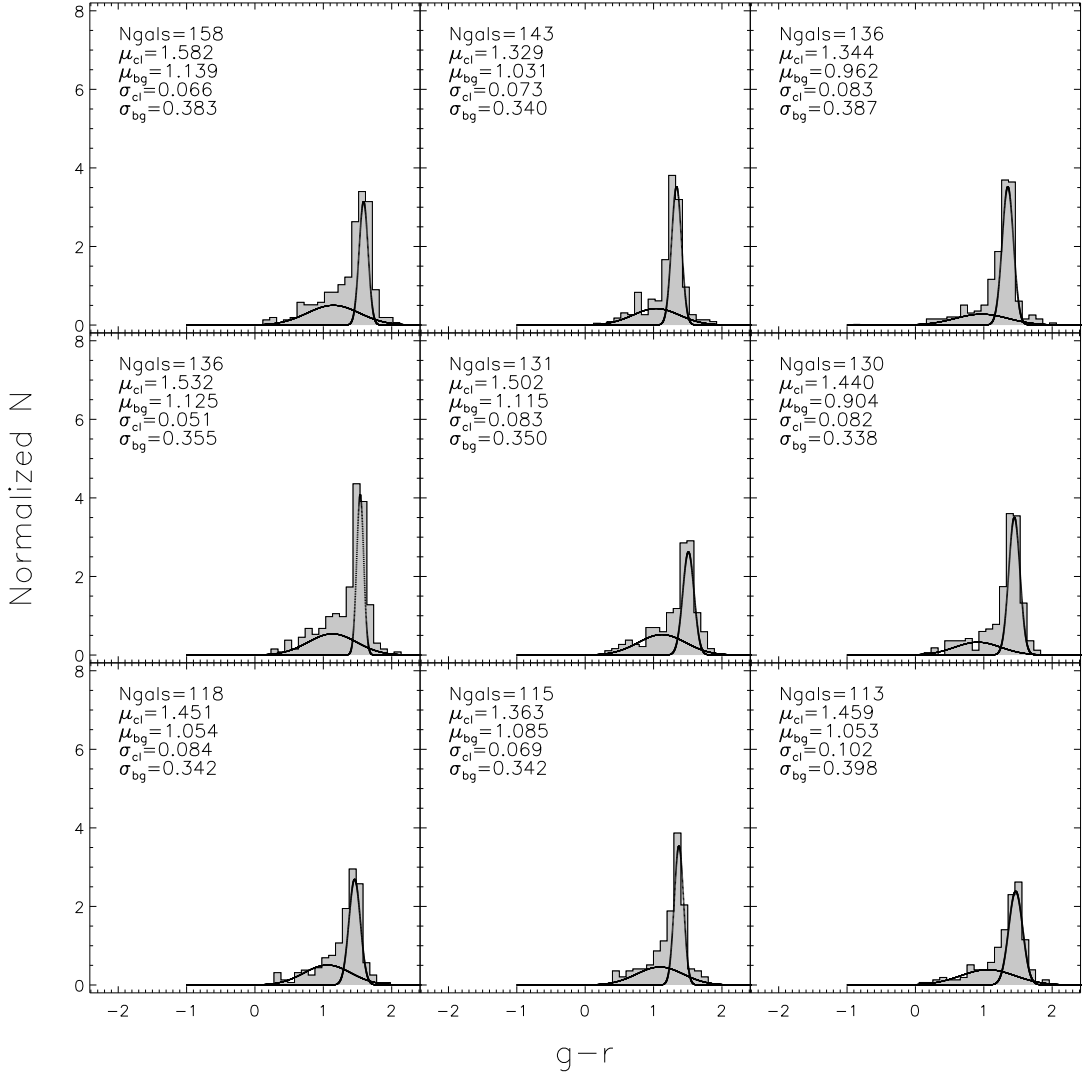} \caption{The ECGMM fitting to the galaxy color 
distribution around 9 rich clusters. Note the fact that we corrected for measurement errors, the two Gaussians appear to be narrower than the histogram.}
\label{fig:bimodal_hist_big}
\end{figure}

\begin{figure}
\epsscale{.9}
\plotone{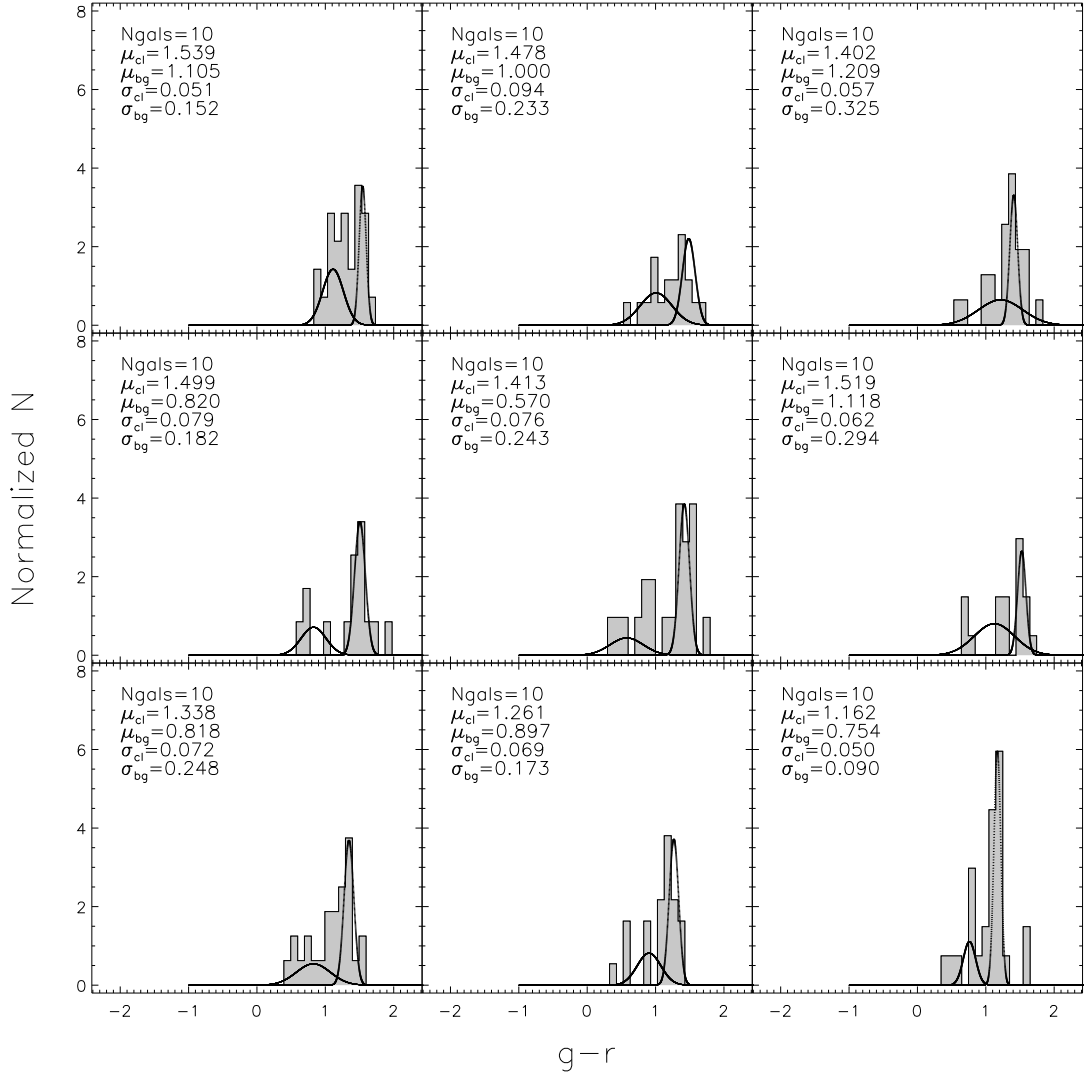} \caption{The ECGMM fitting to the galaxy color 
distribution around 9 small clusters. Note that when there are fewer galaxies histogram is no longer a good way to show the distribution.}
\label{fig:bimodal_hist_small}
\end{figure}

\begin{figure}
\epsscale{.9}
\plotone{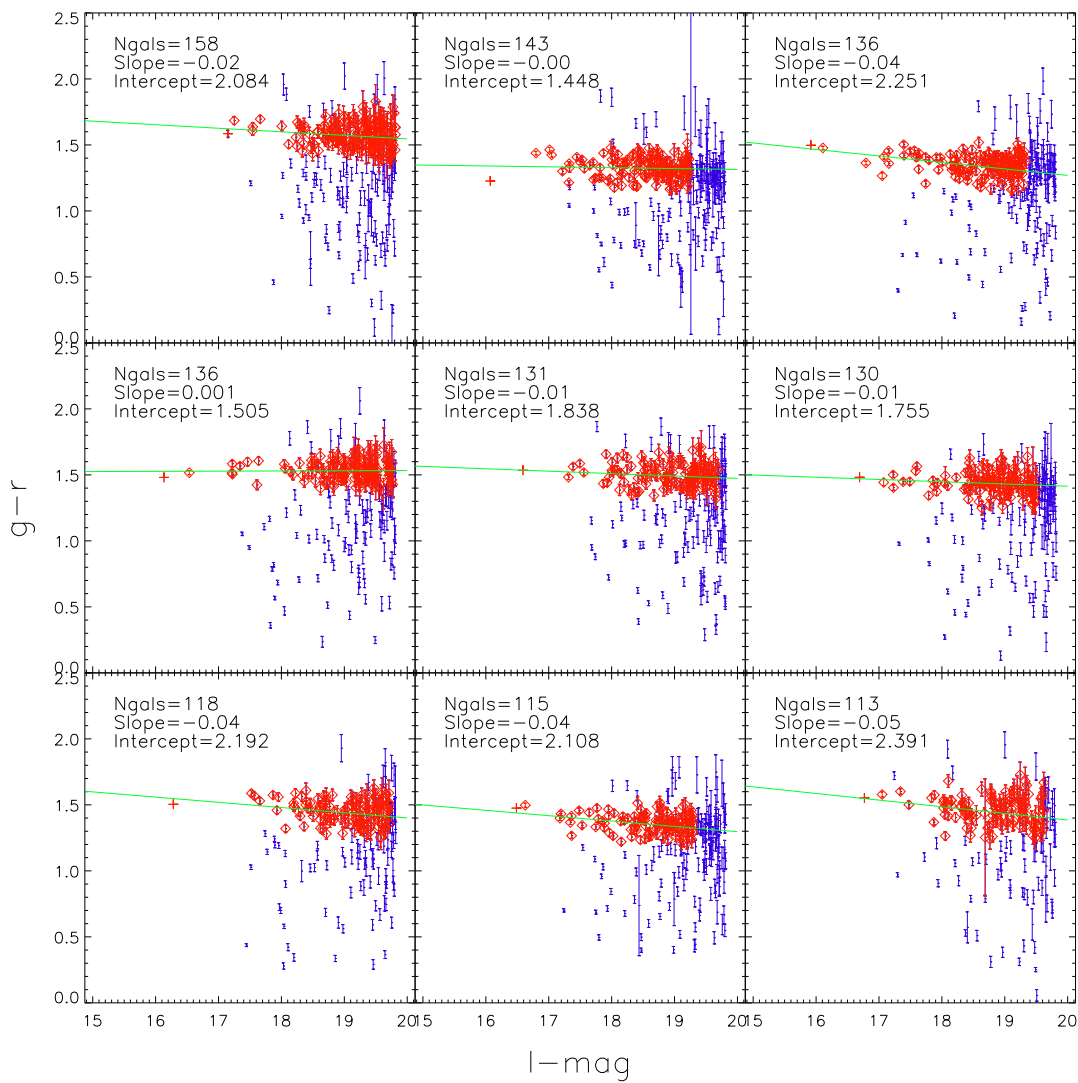} \caption{The CMR around the 9 rich clusters in Fig.\ref{fig:bimodal_hist_big}. The red diamonds and error bars are those from the selected members (red sequence) and the blue dots and error bars are those from field galaxies. The red cross symbol represent the BCG of that cluster. All the galaxies are within $R^{lens}_{200}$ around the BCG, fainter than BCG but brighter than the 0.4L*(see text). The green line is a weighted least square fit (weighted by the inverse square of color errors) to the cluster galaxies.}
\label{fig:bimodal_cmr_big}
\end{figure}

\begin{figure}
\epsscale{.9}
\plotone{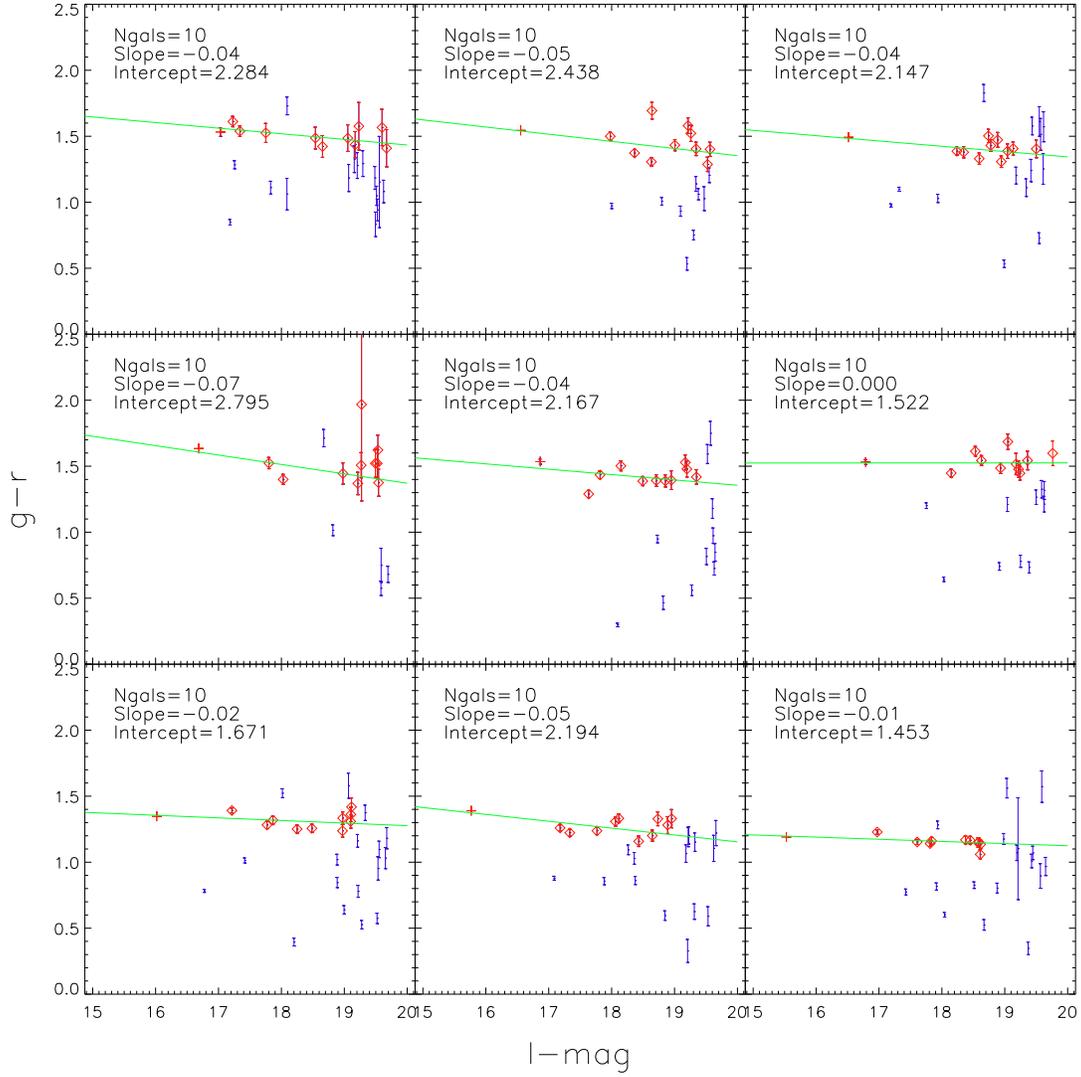} \caption{The CMR around the 9 small clusters in Fig.\ref{fig:bimodal_hist_small}.}
\label{fig:bimodal_cmr_small}
\end{figure}

Next, we identify all SDSS galaxies inside this aperture range,
fainter than the BCG, and brighter than an i band magnitude
corresponding to 0.4 L* at the redshift of the
cluster~\citep{koester07cat}. Then, we apply the ECGMM procedure to
the $g-r$ colors and corresponding measurement errors of these
galaxies. One of the resulting two Gaussian components from the
ECGMM will represent the cluster red sequence color distribution
while the other represents the background/blue galaxy color
distribution. To determine which Gaussian Component belongs to the
cluster, we calculate the likelihood of the BCG's $g-r$ color on
each Gaussian Component. The component for which the BCG has a
higher likelihood is assigned as the cluster component and the other
is declared background. By this way, each maxBCG cluster gets a new richness, $N_{200}^{lens}$. It is worth noting that we apply the above measurements to all maxBCG clusters whose original $N200 \ge 10$. But we will only continue our analysis on a subsample of the clusters whose new measured richness $N_{200}^{lens} \ge 10$ and have two identified Gaussian mixture components in order to guarantee the reliability of our measurements. After this selection, we are left with about 7100 clusters and all our further analysis are based on them. We need to point out that the clusters falling outside of this selection are not necessarily bad clusters. They are just fall below the richness threshold we imposed for quality control. In Fig.\ref{fig:bimodal_hist_big} and Fig.\ref{fig:bimodal_hist_small}, 
we show the ECGMM fitting of 9 big and 9 small clusters as described above. Their corresponding CMRs are plotted in Fig.\ref{fig:bimodal_cmr_big} and Fig.\ref{fig:bimodal_cmr_small}.

\begin{figure}
\epsscale{1.} 
\plotone{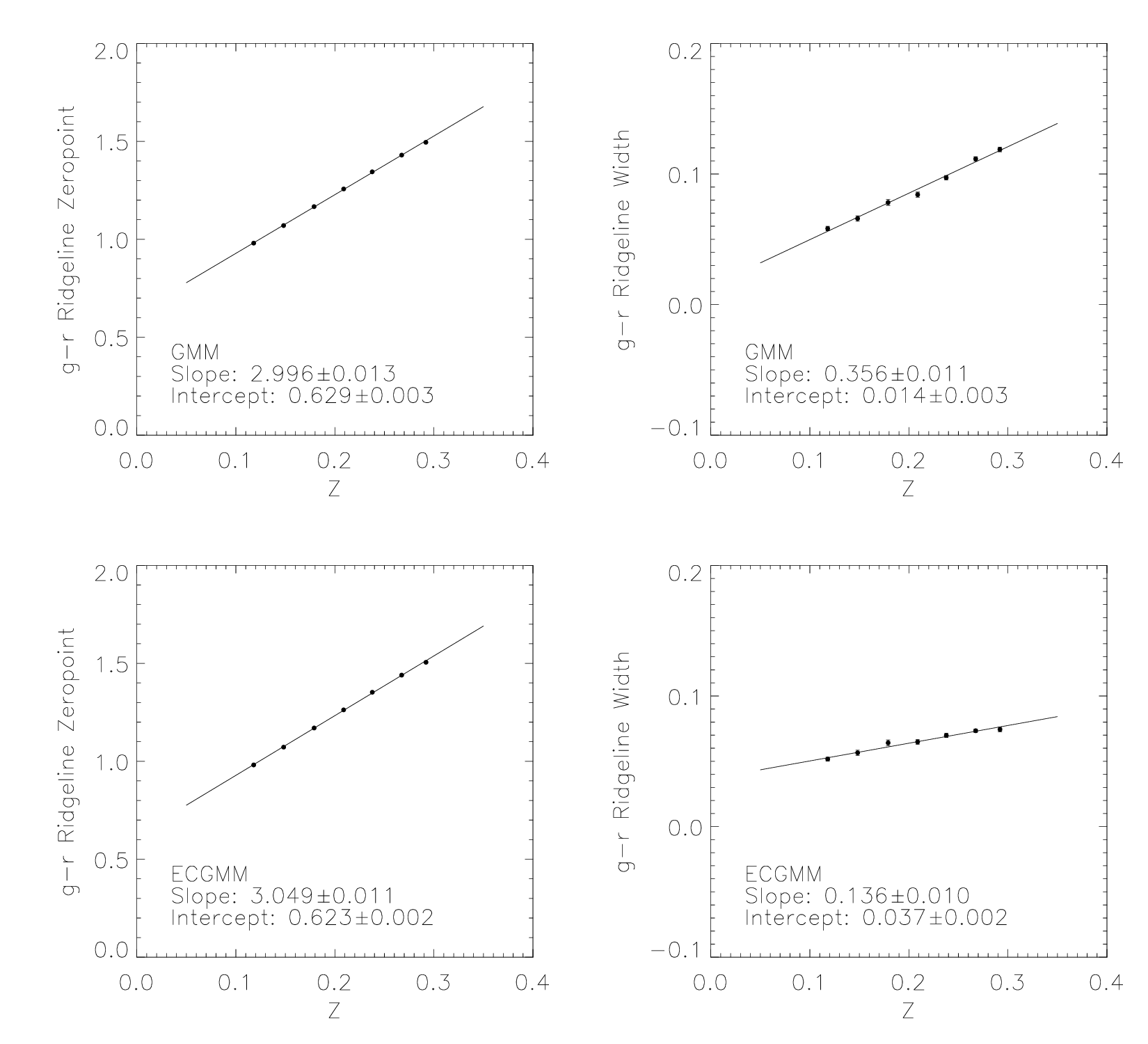} \caption{Tracking the ($g-r$)
red sequence zeropoint and width as a function of redshift, measured
using ordinary GMM (upper panels) and ECGMM (lower panels)
respectively. We bin the measured ridgeline color and width into
redshift bin of 0.04 and then fit the means with a straight line.
After error correction, the broadening of the observed red-sequence
width with redshift is greatly suppressed, revealing the effect of
photometric errors on the observed broadening.}
\label{fig:cl_ridge_z}
\end{figure}

For comparison, we measure the red-sequence location and width using
both ordinary GMM and ECGMM. The top panel of Fig.
\ref{fig:cl_ridge_z} shows the evolution of the average g-r
ridgeline location and width measured using ordinary GMM. We observe
the well-known trend in the average ridgeline zeropoint, and there
is additional apparent strong evolution in the average ridgeline
width, which becomes nearly $140\%$ larger by $z= 0.3$. However,
from the lower two panels which are measured using ECGMM, one can
see very clearly the power of ECGMM in constraining the intrinsic
width of the ridgeline without contamination from measurement error.
The results show that the mean observed $g-r$ ridgeline location
retains the same linear dependence on redshift while the mean scatter
of the ridgeline shows a weak dependence on redshift, with the $g-r$
scatter $\sigma(z=0.1)=0.051 \pm 0.003$ and $\sigma(z=0.3)=0.079 \pm
0.005$ or a broadening by $\sim 55\%$ from $z= 0.1$ to $z= 0.3$.
The strong dependence of the scatter on redshift
from the GMM is mostly due to the increased measurement errors for
cluster members at higher redshift.

\section{The red sequence ridgeline slope}

\subsection{Ridgeline slope from galaxy clusters}

It has been pointed out that the color-magnitude relation (CMR) of
cluster member galaxies has a negative slope
~\citep[e.g][]{kodama97,gladders98}, so that fainter member galaxes
are generally bluer. The evolution of these CMR slopes with respect
to redshift and richness has been difficult to address, largely due
to the lack of a sufficiently large cluster catalog with well
measured photometry for all its galaxies. The maxBCG catalog
provides about 14,000 galaxy clusters, extending over $0.1 \leq z
\leq 0.3$, which enables us to measure the slope of the CMR for
clusters with good statistics across a range in both richness and
redshift.

Measurement of the slope of the CMR typically proceeds by
identification of the cluster red-sequence, followed by some
iterative process of outlier removal, and a determination of cluster
``member'' galaxies which are then used to measure the slope and
zeropoint of the CMR.\footnote{In \citet{Andreon06}, a slightly different method was introduced by directly modeling the CMR and measurement errors into the likelihood function without separating the red sequence galaxies first.} We apply the method described in previous
sections to measure the color distribution of individual clusters
and to assign the memberships for every cluster by requiring the
color difference between the member galaxies and ridgeline within 
$\pm 2\sigma$ ($\sigma$ is the convolved ridgeline width,
given by the best-fit ECGMM, and the measurement errors of individual
member galaxy's color). The richness of the cluster measured by this
way is denoted as $N_{200}^{lens}$. We choose $2\sigma$ because this
is roughly where the background component's likelihood dominates
over the cluster component's likelihood. Based on this
identification of membership driven by the ECGMM, we fit for the CMR of clusters 
galaxies with a straight line using weighted least square fitting. The weights we used are the inverse square of the measurement errors of $g-r$. We call the slope of the fitted line as the slope fo the ridgeline.

\begin{figure}
\epsscale{0.9} \plotone{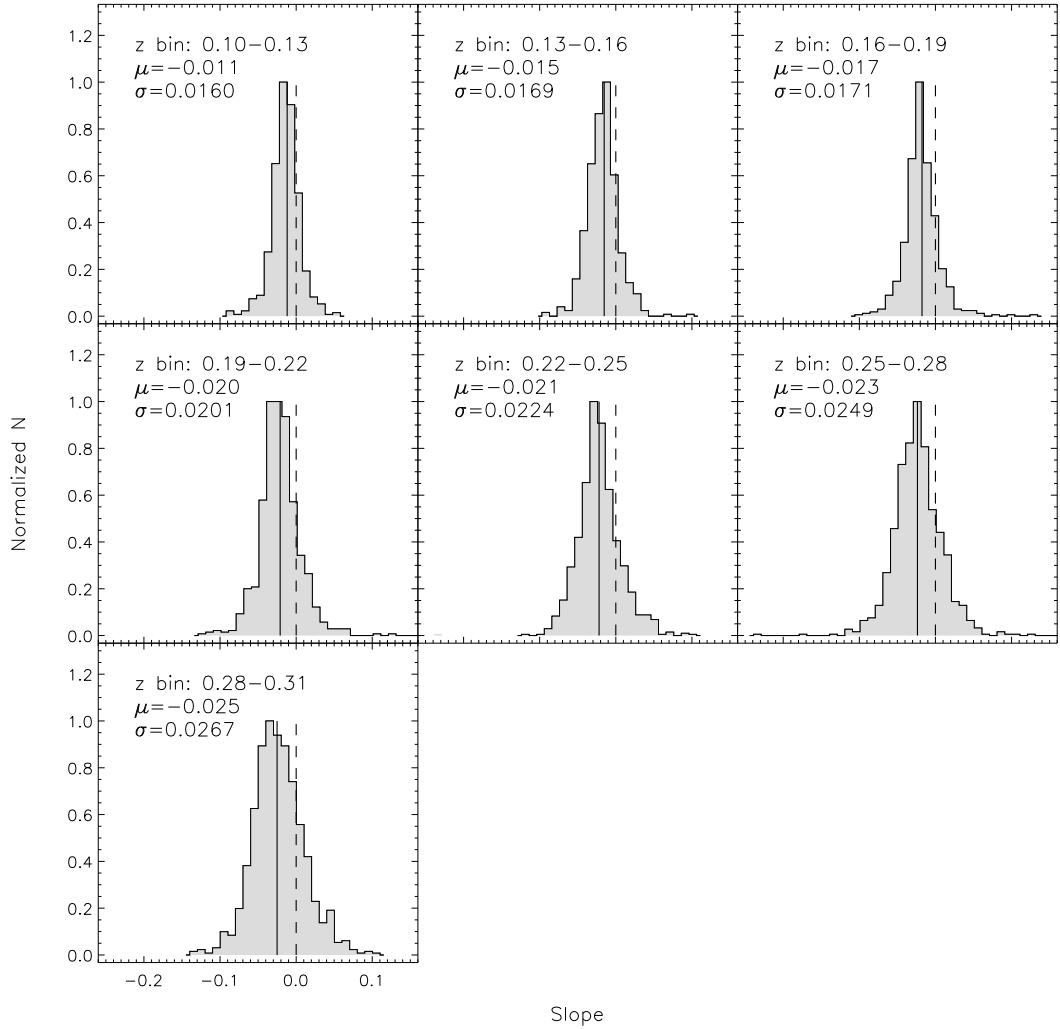} \caption{The distributions of
measured ridgeline slopes for clusters in steps of 0.03 in redshift.
$\mu$ and $\sigma$ denote the mean and width of the distribution. The
dashed line corresponds to zero.} \label{fig:zbinslope}
\end{figure}

\begin{figure}[tb]
\epsscale{0.7} \plotone{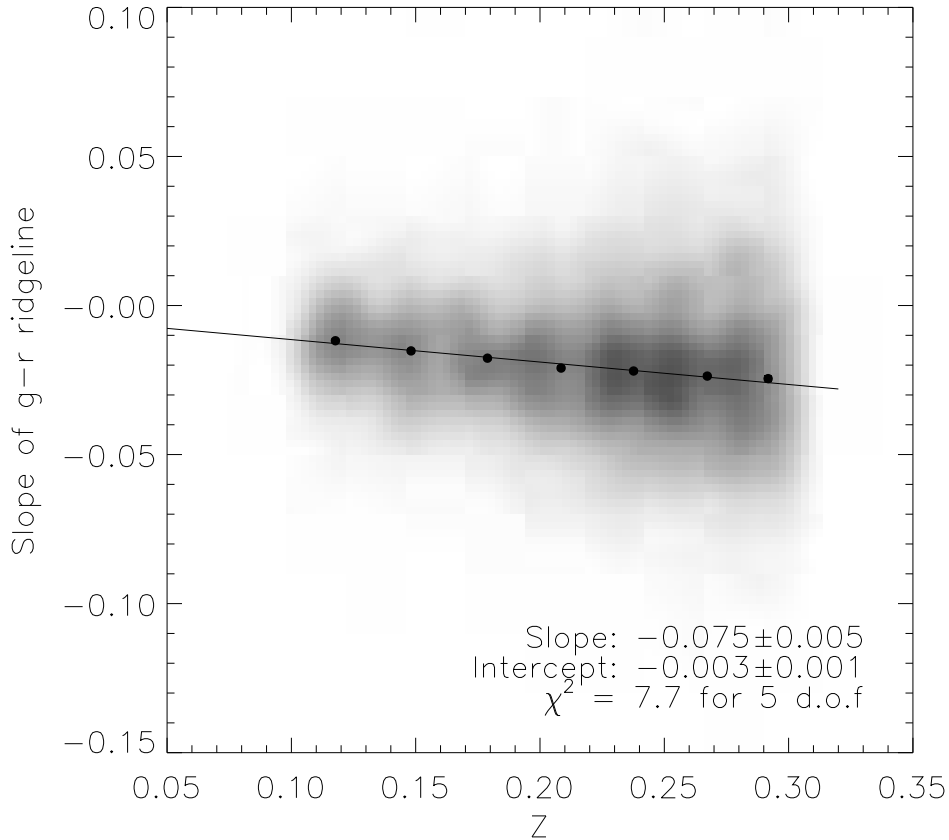} \caption{Tracking the
observed red-sequence slope vs redshift. The gray clouds represent
the slope measurements from individual clusters. The black solid
circles and error bars are weighted mean and the standard deviation
to the weighted mean for each redshift bin ($\Delta=0.03$). Note
that the error bars in the plot are smaller than the symbols.}
\label{fig:slope_ztrend}
\end{figure}

\begin{figure*}[tb]
\epsscale{0.9} \plotone{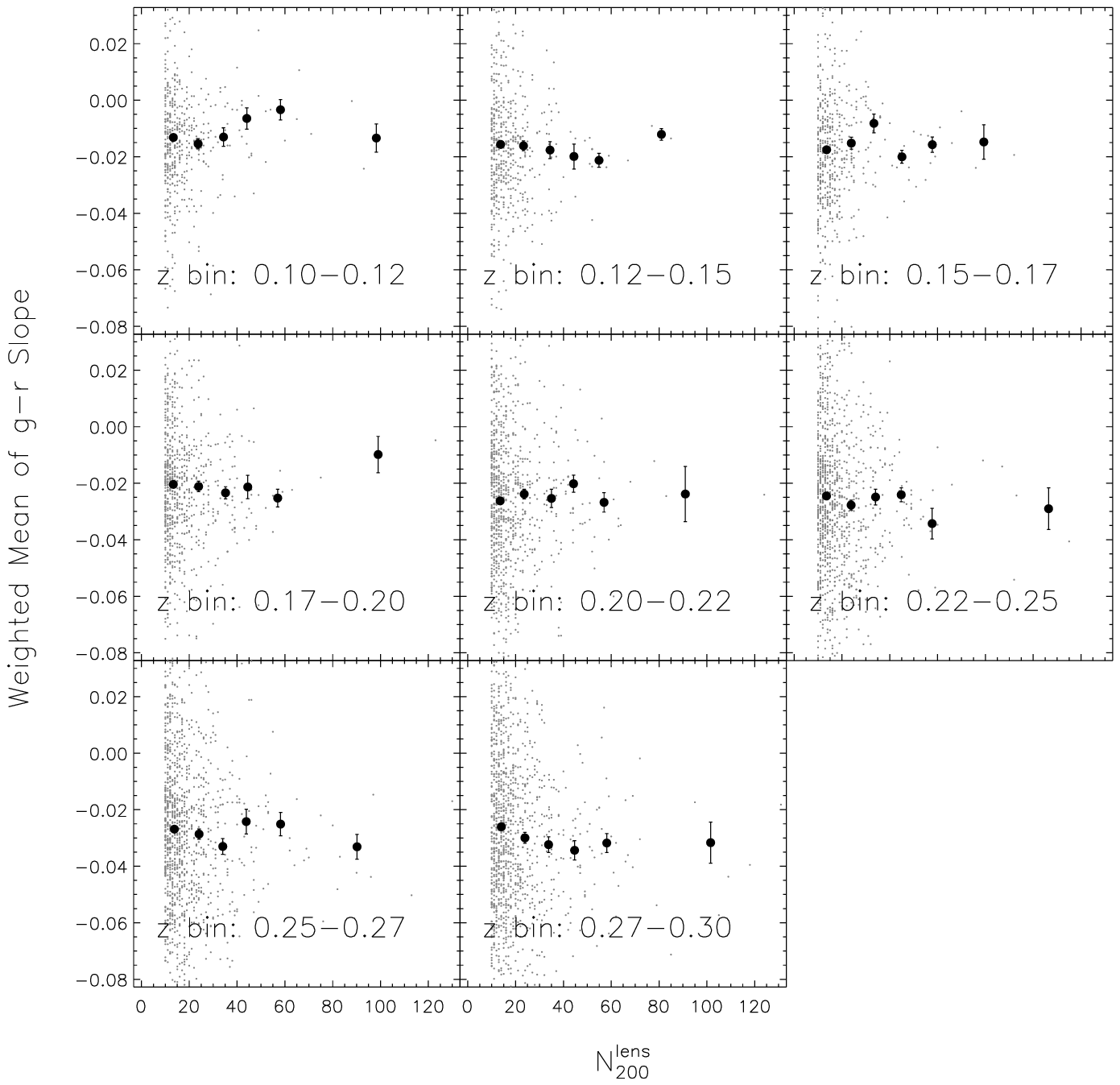} \caption{The evolution
of mean ridgeline slope vs richness at different redshift slices.
The richness bins brackets are chosen as
$N_{200}^{lens}=$[10,20,30,40,60,80,161]. The light dark points are
from individual clusters. The black solid dots and error bars are
weighted mean and standard deviation of the slopes in each
$N_{gals}$ bin for every redshift slice. From the plot, we did not
see strong trends of the slope evolution w.r.t richness.}
\label{fig:zbin_ngals_trend}
\end{figure*}

The distribution of ridgeline slopes for maxBCG clusters is shown
in Fig.\ref{fig:zbinslope} in bins of $\Delta z=0.03$. Despite the
substantial scatter in slope among individual clusters, we can see
from Fig.\ref{fig:zbinslope} and Fig.\ref{fig:slope_ztrend} that the
mean slope of the red sequence ridgelines for clusters deviates from
zero for $0.1 \leq z \leq 0.3$. For any bin, the error on the mean
places the measurement many standard deviations from zero.

In Fig.\ref{fig:slope_ztrend}, it is apparent that the observed
trend of the mean ridgeline slope with redshift is statistically
significant: the slope becomes steeper by a factor of 2.5 by
$z=0.3$. In Fig.\ref{fig:zbin_ngals_trend}, we plot the evolution of
the slopes vs richness in each redshift slice, which shows that the
dependencies of the ridgeline slope with respect to richness is weak, as
shown elsewhere~\citep[e.g.][]{hogg04}. Clearly, the observed slope
of the red-sequence is not associated with cluster richness, and is
unsurprisingly a strong function of redshift (see Discussion).

\subsection{Ridgeline slope from spectroscopic data}\label{spec_tilt}

The above measurement is based only on a photometric determination
of red sequence galaxies. The level to which projection plays into
this selection is as yet unknown. The true red-sequence galaxy population
in some physical volume, either in a cluster or in the field, is contaminated
by dusty foreground galaxies which can be rejected via spectroscopy, and
by the peculiar velocities of the galaxies themselves.

To address the possibility of foreground contamination, it is
interesting to see if the above results are preserved in a
spectroscopic sample of galaxies. To achieve this goal, we use
galaxies with spectra from DR6 of SDSS Value Added Galaxy
Catalog~\citep[VAGC]{blantonvagc} for a comparison. Due to the
selection effects of the spectroscopic data, we will choose only the
galaxies in redshift from 0.1 to 0.2 and brighter than 0.4 L*
magnitude at their respective redshifts. By extension from the photometric sample and from previous
work~\citep{hogg04}, we know that the slope does not vary with
environment, so the field sample represented by our spectroscopy
should be a fair representation of the expected slope in clusters.

\begin{figure*}[tb]
\epsscale{0.8} \plotone{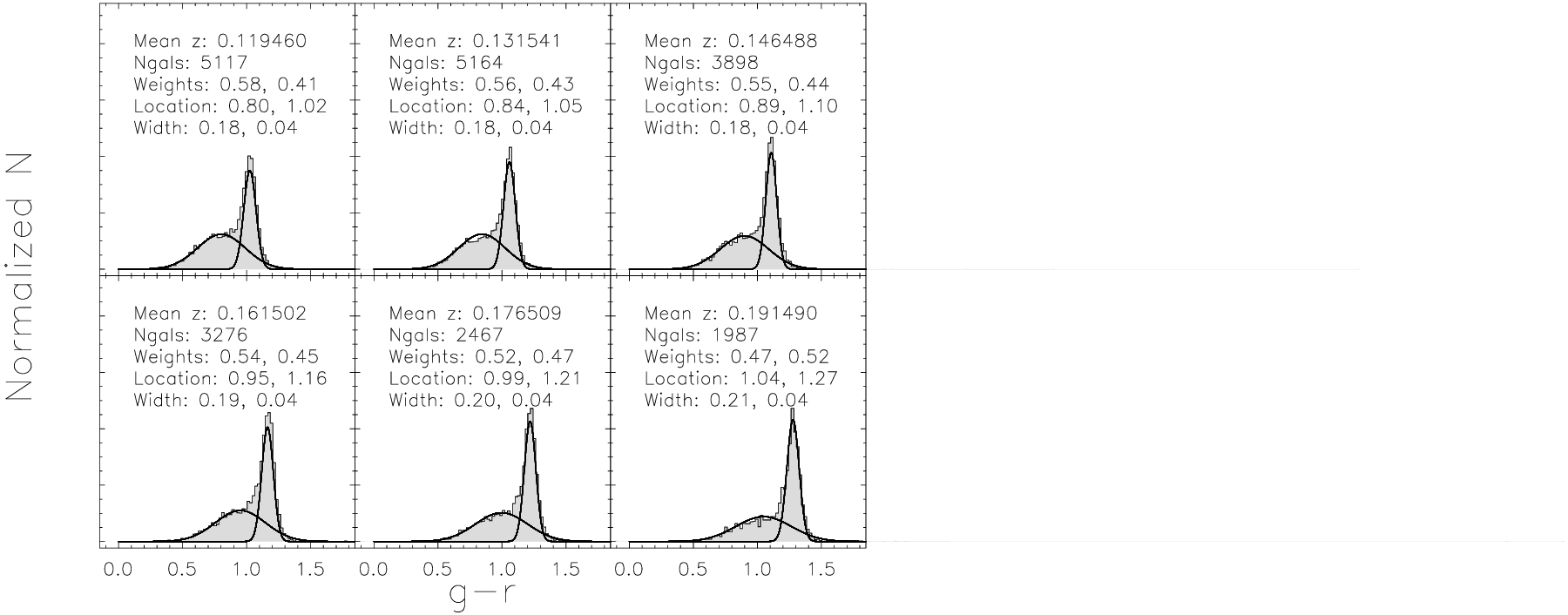}
\plotone{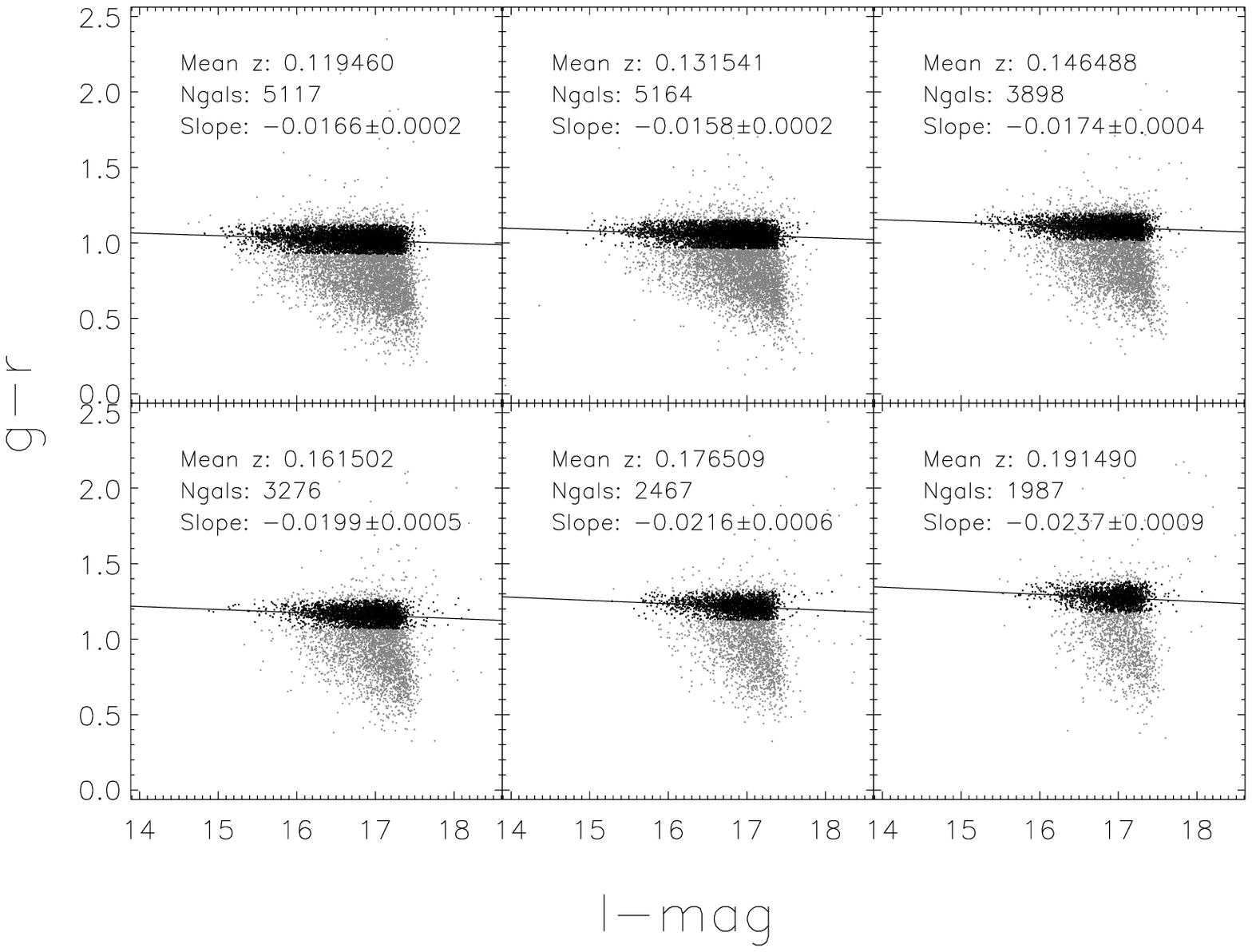} \caption{Evaluating the ECGMM-derived
red sequence slopes in SDSS spectroscopy of field galaxies. The
normalized color histograms (top panel) for $\Delta z =0.003$ slices
in spectroscopic redshift clearly show the presence of the red and
blue components in the field galaxy distribution. ECGMM is used to
separate the two components, the redder of which is to measure the
CMR (bottom panel).} \label{fig:spec_red_blue_hist_cmd}
\end{figure*}

\begin{figure}
\epsscale{0.9} \plotone{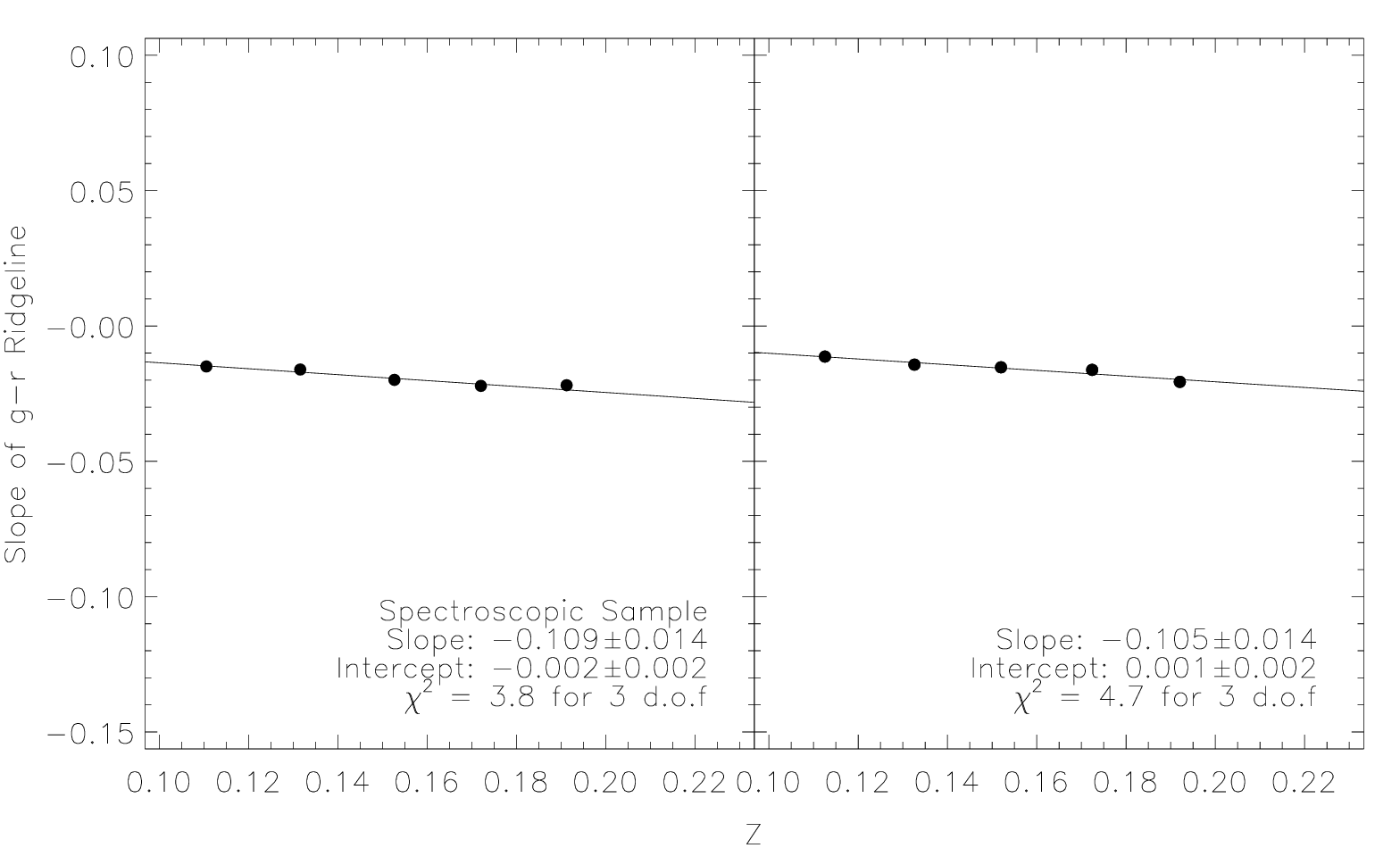}
 \caption{The comparison of the evolution of the slopes of CMR for spectroscopic
 sample and cluster sample. Since the spectroscopic sample is biased
 due to selection effects at $z \ge 0.2$, we choose both sample in the redshift
 range from 0.1 to 0.2 and then bin the slopes in redshift bin of 0.02.
 We then fit straight line to the mean slopes in each bin for spectroscopic sample and cluster sample respectively.}
\label{fig:spec_cluster_z}
\end{figure}

Our procedures are as follows: we first bin the galaxies into bins
of size $\Delta z = 0.003$, which corresponds to velocity slices of
900km/sec. The color distribution of the galaxies in each bin shows
clear bimodality (top panel of Fig.
\ref{fig:spec_red_blue_hist_cmd}). Then, we separate the red
sequence galaxies in each bin using ECGMM. The red sequence galaxies
correspond to the Gaussian component with bigger $g-r$ value and we
choose $\pm 2\sigma$ from the peak location as red galaxy samples
for each redshift slice, in a fashion similar to the one we used for
cluster galaxies. Then, in every bin, we fit the CMR of galaxies' $g-r$ colors and
$i$-band magnitude with a straight line using weighted least square fitting with the weights come from the inverse sequare of the color measurement errors. The weights are the inverse square of the $g-r$ measurement errors. We record the corresponding slopes. In the bottom panel
Fig.\ref{fig:spec_red_blue_hist_cmd}, we choose 6 redshift bins
($\Delta z = 0.003$) to illustrate the red/blue galaxy separation
and the ridgeline slope fitting in each bin. Finally, we fit the
variation of slope with redshift with a line to look for a trend,
the results are shown in the left panel of
Fig.\ref{fig:spec_cluster_z}. As a comparison to the cluster sample,
we also plot the mean variation of the ridgeline slope from clusters
in the same redshift range [0.1, 0.2] in the right panel of
Fig.\ref{fig:spec_cluster_z}. When the redshift range is changed,
the slope of the fitted line for the cluster sample becomes steeper
as compared to Fig.\ref{fig:slope_ztrend}. The reason lies in that
the linear fit to the trend is only the first order approximation.
But for our purpose here, we just need to require the cluster sample
and spectroscopic sample on the same redshift range so as to compare
them fairly.

Comparing the two panels in Fig.\ref{fig:spec_cluster_z} shows that
the two slopes from spectroscopic sample and cluster sample are not
different in a statistically significant way. This further confirms
our previous observation that the cluster environment will not
affect the slope.

\section{Discussion}
To this point, the measurements have been presented in the observed
frame. Photometric cluster detection and the quantities derived
(e.g. richness) operate in the frame of the observer, and
predictions from galaxy formation models and mock galaxy catalogs
can be evaluated in light of these precision measurements. They have
particular applicability to calibration of optical cluster detection
efforts, especially to those that rely on the properties of the red
sequence. The methodologies developed herein allow the
``bootstrapping'' of optical algorithms: basic cluster finders
locate the clusters, and precision measurements (such as these) of
said clusters lead to refinements in those algorithms. The extent to
which each of these agrees with previous measurements from
spectroscopy, other cluster samples, and simulations is left for
future work. However, for illustrative purposes, we list the
relevant observational considerations to be made in understanding
the context of these measurements with respect to previous work in
the literature, and then highlight a few of our more interesting
results.

In general, there are five places where the comparison to previous
work must be treated with caution, which can be summarized as
follows: 1) redshifting of the galaxy spectra through the bandpasses
under consideration, which imparts trends in the observed colors, 2)
selection effects imposed by the color selection
\citep[e.g.][]{franzetti07}, 3) aperture effects, i.e. the aperture
used to measure the color in different banspasses
~\citep[e.g.][]{scodeggio01,blakeslee06}, 4) projection effects. 5)
actual evolution in the red sequence.

To some level, any of the aforementioned issues may play into our
results: (i) at $z \simeq 0.1$, the CMR of photometrically-selected
galaxies is noticeably shallower than previous spectroscopic
measurements of the color magnitude relation~\citep{hogg04,cool06},
(ii) the slopes are almost independent of cluster richness; (iii)
the photometric error-corrected scatter of the red-sequence broadens
mildly with redshift; (iv) the observed mean slope of the CMR is
negative and it becomes more negative as redshift increases.

Naively, we expect that our measurement of the slope of the
red-sequence, $-0.013 \pm 0.0003$ mags mag$^{-1}$ at $z=0.1$, 
corresponds to the SDSS spectroscopic analysis of~\citet{hogg04},
for which the slope is -0.022 mags mag$^{-1}$ in $^{0.1}(g-r)$. In
addition to the fact that the~\citet{hogg04} measurements are
k-corrected to the $z=0.1$ rest-frame, one possible difference comes
from our definition of the red sequence:~\citet{hogg04} use a
$2\sigma$ clipping algorithm to define the red sequence and to
iteratively reject outliers. While they split the sample by Sersic
index, sigma-clipping may be more permissive of objects near the
``blue cloud'' to be included in the red-sequence, while the method
presented in this paper automatically accounts for the presence of
these objects. Our slope measurements at a given redshift may also
be biased shallow, as the initial 2 $\sigma$ cut derived from the
ECGMM fit does not account for the slope in the red-sequence itself,
i.e. the cut is applied in the same way regardless of magnitude.
Ideally, an iterative procedure would be employed to determine the
best-fit line for each cluster and the $2 \sigma$ cut would be
applied as a function of magnitude. Unfortunately, the small number
statistics for low richness clusters do not permit this to be
implemented in a robust fashion.

Insofar as richness and local density are similar indicators of
environment, the second observation (ii) that the slope is almost
independent of environment is in basic agreement with
~\citet{hogg04}, who use SDSS spectroscopy at $z\sim 0.1$ to compare
galaxies with high ($n \ge 2$) Sersic indicies in different
environments characterized by their local density.

After the photometric error correction performed by ECGMM, a trend
in the scatter with redshift remains (iii), such that the scatter
increases with increasing redshift. At high redshift $z \simeq 1$,
the color-magnitude relation has been measured in a handful of
clusters~\citep[e.g.]{mei09,koester09,santos09} with the general
conclusion that the restframe scatter in the CMR does not evolve
with redshift. More locally, the SDSS Luminous Red Galaxy (LRG)
Sample has been used to measure various redshifted frames of bright
($L \gtrsim 2.2L_*$) red galaxies~\citep{cool06}.~\citet{cool06} 
find the intrinsic rest-frame scatter $^{0.16}(g-r)=35.4 \pm 3.7$
and $^{0.37}(g-r)=43.5 \pm 6.2$ mmags, consistent with no evolution. 
However, with increased cluster sample, our $observed$ frame measurements reveal an increase in the scatter, 
shown by a statistically significant non-zero slope (the bottom right panel in Fig.~\ref{fig:cl_ridge_z}).



%

Result (iv) is in qualitative agreement with the results in
~\citep{gladders98} who find a similar trend in the slope for a
sample 44 Abell clusters at $z \le 0.15$ and 6 clusters at $0.2 \le
z \le 0.75$, the largest previous study of its kind. In their study
of the scatter of the CMR in LRGs,~\citet{cool06} report no
significant trend with redshift in the rest-frame slope of LRGs over
$0.16 < z < 0.37$ in either the cluster or the field, but caution
that the sample is not-well suited to measuring the slope. The
observed factor of 2.5 increase in the magnitude in our measurement
of the slope is likely due to a combination of the lack of
k-corrections and selection effects~\citep[e.g.][]{franzetti07}
derived from color cuts that may preferentially include a larger and
larger fraction of galaxies with significant star-formation at
increasing redshifts.

A further contribution to the inflated slope may come from the
choice of the color aperture.~\citet{vandokkum98} and
~\citet{scodeggio01} note the importance of the use of adaptive
apertures, which place the color measurements of large and small
galaxies on the same footing. This point motivates our choice of
$\tt{MODEL\_MAGS}$ from the SDSS, which are derived from the
best-fit convolution of the local PSF with a deVaucoleurs model in
the $r$-band. This same best-fit model is then used to compute the
flux in both the $g$ and $r$-bands.

As our results indicated, the intrinsic ridgeline slope may decrease as redshift increase. Though there are many factors that may complicate the interpretation of this results as we discussed in the preceding paragraphs, it is still worth to speculate what may lead to the intrinsic evolution. In general, galaxies are more active at higher redshift~\citep{cowie96}, and their color distributon spreads wider and toward bluer end. Therefore, clusters at higher redshift are more likely to have member galaxies with wider and bluer color distribution. As a result, when we fit the CMR with a straightline, we tend to have more negative slopes at higher redshift.


\section{Summary}

In this paper, we have presented the ECGMM, a new purely photometric
method which characterizes the red sequence ridgeline in cluster
samples with large statistics. This provides precise measures of the
mean variation of the red sequence ridgeline location and scatter
(width) with respect to redshift, properly corrected for photometric
errors. The measured slopes, scatters, and zeropoints are directly
applicable to improved cluster finding efforts and to
characterization of known galaxy clusters.

Applying the method to maxBCG clusters approximately recovers known
properties of the red sequence, namely its slope and the variation
of the slope with redshift, and the insensitivity of the slope to
environment. It also suggests that the scatter of the red-sequence
increases mildly with redshift, and that the slope of the
red-sequence grows substantially by $z \simeq 0.3$, but we caution
that these observed trends may be attributable to a host of
observational effects that we have made no attempt to correct. Color
selection effects, the lack of k-corrections, and the details of the
measurement of the individual cluster CMRs require proper attention
before applying these results to models of galaxy formation.
Nonetheless, these measurements can serve as an important
observational check on simulation and mock galaxy catalogs.


\section*{Acknowledgments}

JH thank Hyunsook Lee for helpful conversation about the BIC and
Mixture models. JH and TM gratefully acknowledge support from NSF
grant AST 0807304 and DoE Grant DE-FG02-95ER40899. ESR would like to
thank the TABASGO foundation. ER was funded by the Center for
Cosmology and Astro-Particle Physics at The Ohio State University
and by NSF grant AST 0707985. This project was made possible by
workshop support from the Michigan Center for Theoretical Physics.

Funding for the creation and distribution of the SDSS Archive has
been provided by the Alfred P. Sloan Foundation, the Participating
Institutions, the National Aeronautics and Space Administration, the
National Science Foundation, the U.S. Department of Energy, the
Japanese Monbukagakusho, and the Max Planck Society. The SDSS Web
site is http://www.sdss.org/. The SDSS is managed by the
Astrophysical Research Consortium (ARC) for the Participating
Institutions. The Participating Institutions are The University of
Chicago, Fermilab, the Institute for Advanced Study, the Japan
Participation Group, The Johns Hopkins University, the Korean
Scientist Group, Los Alamos National Laboratory, the Max-Planck-
Institute for Astronomy (MPIA), theMax-Planck-Institute for
Astrophysics (MPA), New Mexico State University, University of
Pittsburgh, University of Portsmouth, Princeton University, the
United States Naval Observatory, and the University of Washington

\appendix{

\section{The recursive relation for the error corrected Gaussian Mixture Model}

In this appendix, we show the derivation of the likelihood function
Eq.~\ref{totallk} and the EM recursive relations for the error
corrected GMM. To begin with, we introduce the following notations
in Table.\ref{table:notations}. For brevity, we denote the
parameters ($\mu_i$, $\sigma_i$ and $w_i$) collectively by $\theta$
and $(t)$ represents the $t^{th}$ iteration. $M$ represents number
of data points and $N$ represents the number of mixtures.

\begin{table}
\begin{center}
\scriptsize
\begin{tabular}{l l}
\hline\hline Notations & Meaning \\
\hline
$y_1$,$...$, $y_j$, $...$, $y_m$: & Observed colors of BCGs and member galaxies.\\
$\bar{y}_1$,$...$, $\bar{y}_j$, $...$, $\bar{y}_m$: & True colors of BCGs and member galaxies.\\
$z_1$,$...$, $z_j$, $...$, $z_m$: & Hidden variables that tell which Gaussian component the $\bar{y}_j$ is sampled from.\\
$\delta_1$,$...$, $\delta_j$,$...$, $\delta_m$: & Measurement errors for every $y_j$.\\
$\mu_1$, $...$, $\mu_i$, $...$, $\mu_n$: & Mean of each Gaussian component.\\
$\sigma_1$, $...$, $\sigma_i$, $...$, $\sigma_n$: & Width of each Gaussian component.\\
$w_1$, $...$, $w_i$, $...$, $w_n$: & Weights of corresponding
Gaussian components.\\ \hline
\end{tabular}
\caption{The notations used in our derivation of ECGMM algorithm}
\label{table:notations}
\end{center}
\end{table}

Since we assume the true color distribution can be approximated by
mixture of Gaussian distributions, we have the following probability
density function for $p(\bar{y_j}|\theta)$:
\begin{equation}\label{truep}
p(\bar{y_j}|\theta)=\sum_{i=1}^{N}\frac{1}{\sqrt{2 \pi \sigma_i^2}}
\exp\bigg[-\frac{(\bar{y}_j-\mu_i)^2}{2\sigma_i^{2}}\bigg]
\end{equation}

\noindent Though the true colors are not directly observable, we
know that its distribution given the observed colors and measurement
errors is approximately Gaussian:

\begin{equation}\label{ppri}
p(\bar{y_j}|y_j)=\frac{1}{\sqrt{2 \pi \delta_j^2}}
\exp\bigg[-\frac{(\bar{y}_j-y_j)^2}{2\delta_j^{2}}\bigg]
\end{equation}

\noindent then, the likelihood function (under the flat priors for
$\theta$)
\begin{equation}\label{singlel}
L(\theta|y_j)=\int p(\bar{y_j}|\theta)p(\bar{y_j}|y_j)d\bar{y_j}
\end{equation}

\noindent After integrating over $\bar{y}_j$ and extending to all
data points ($\prod_{j=1}^M$), we arrive at Eq.\ref{totallk}. The
optimal parameters could be obtained by maximizing the above
likelihood. However, if we introduce hidden variables, $z$, that
tell us which Gaussian component the $y_j$ is sampled from, then the
whole maximization process could be significantly simplified. The
corresponding pdf of data given $z$ and $\theta$ is

\begin{equation}\label{knowpdf}
p(y|z_j=i,\theta^{(t)})=\prod_{j=1}^M
p(y_j|z_j=i,\theta_i^{(t)})=\prod_{j=1}^M \frac{1}{\sqrt{2 \pi
(\sigma_i^{(t)2}+\delta_j^{2})}}\exp\bigg[-\frac{(y_j-\mu_i^{(t)})^2}{2(\sigma_i^{(t)2}+\delta_j^{2})}\bigg]
\end{equation}

\noindent We use $w_i$ denote the weight of each Gaussian Component
in the mixture and is given by $w_i=p(z_j=i|\theta)$. The estimation
of hidden variable could be related to Eq.\ref{knowpdf} by Bayes'
Theorem as following:

\begin{equation}
p(z_j=i|y_j,\theta^{(t)})=\frac{p(z_j=i,y_j|\theta^{(t)})}{p(y_j|\theta^{(t)})}=\frac{p(y_j|z_j=i,\theta^{(t)})p(z_j=i|\theta^{(t)})}{\sum_{i=1}^N p(y_j|z_j=i,\theta^{(t)})p(z_j=i|\theta^{(t)})}
\end{equation}
The EM algorithm iteratively update the parameters $\theta$ by
maximizing the expected log likelihood

\begin{equation}
Q(\theta)=\sum_{i=1}^{N}\sum_{j=1}^{M}p(z_j=i|y_j,\theta^{(t)})\bigg[-\frac{1}{2}\ln(2
\pi)-\frac{1}{2}\ln(\sigma_i^2+\delta_j^2)-\frac{(y_j-\mu_i)^2}{2(\sigma_i^2+\delta_j^2)}+\ln
p(z_j=i|\theta^{(t)})\bigg]
\end{equation}

\noindent under the constraint $\sum_{i=1}^N
p(z_j=i|\theta^{(t)})=1$. Using the Lagrange Multiplier approach,
we redefine

\begin{equation}
\tilde Q(\theta)=Q(\theta)-\lambda \bigg[\sum_{i=1}^N
p(z_j=i|\theta^{(t)}) -1\bigg]
\end{equation}

\noindent with $\lambda$ as the multiplier.

\begin{equation}\label{partialmu}
\frac{\partial \tilde Q(\theta)}{\partial \mu_i}=
\sum_{j=1}^M\bigg[p(z_j=i|y_j,\theta^{(t)})\bigg(\frac{y_j-\mu_i}{\sigma_i^2+\delta_j^2}\bigg)
\bigg]=0
\end{equation}

\noindent From Eq.\ref{partialmu}, we can arrive at the following
recursive relation for $\mu$:

\begin{equation}\label{gmm1}
\mu_i^{(t+1)}=\frac{\sum_{j=1}^{M}y_jp(z_j=i|y_j,\theta_i^{(t)})/(1+\delta_j^{2}/\sigma_i^{(t)2})}{\sum_{j=1}^{M}p(z_j=i|y_j,\theta_i^{(t)})/(1+\delta_j^{2}/\sigma_i^{(t)2})}
\end{equation}

\noindent Similarly, we have

\begin{equation}\label{partialsigma}
\frac{\partial \tilde Q(\theta)}{\partial \sigma_i}= \sum_{j=1}^M
p(z_j=i|y_j,\theta^{(t)})\bigg[\frac{\sigma_i^2(1+\delta_j^2/\sigma_i^2)-(y_j-\mu_i)^2}{\sigma_i^4(1+\delta_j^2/\sigma_i^2)^2}\bigg]
=0
\end{equation}

\noindent Note that since $\sigma_i$ and $\delta_j$ are entangled
within the summation, there would not be an simple analytic solution
for $\sigma_i$. However, since the algorithm is iterative in nature
and the major contribution for the update of $\sigma_i$ is from
$(y_j-\mu_i)^2$, we could approximate $\sigma_i$ in
$\delta_j^2/\sigma_i^2$ with its value in $t^{th}$ iteration. Then
we can solve for the $(t+1)^{th}$ iteration relation for $\sigma_i$
as:

\begin{equation}\label{gmm2}
\sigma_i^{(t+1)}=\bigg[\frac{\sum_{j=1}^{M}(y_j-\mu_i)^2p(z_j=i|y_j,\theta_i^{(t)})/(1+\delta_j^{2}/\sigma_i^{(t)2})^2}{\sum_{j=1}^{M}p(z_j=i|y_j,\theta_i^{(t)})/(1+\delta_j^{2}/\sigma_i^{(t)2})}\bigg]^{1/2}
\end{equation}

\noindent Our numerical test shows that such an approximation works
fine in practice. For $w_i=p(z_j=i|\theta)$, we have

\begin{equation}\label{partialwi}
\frac{\partial \tilde Q(\theta)}{\partial w_i}= \sum_{j=1}^M
p(z_j=i|y_j,\theta^{(t)})/w_i -\lambda=0
\end{equation}

\noindent which leads to

\begin{equation}\label{wi}
w_i=p(z_j=i|\theta)=\frac{1}{\lambda}\sum_{j=1}^M
p(z_j=i|y_j,\theta^{(t)})
\end{equation}

\noindent Using the condition $\sum w_i=1$, we have $\lambda=M$.
Substitute $\lambda$ back to Eq. \ref{wi}, we arrive at:

\begin{equation}\label{gmm3}
w_i^{(t+1)}=\frac{1}{M}\sum_{j=1}^{M}p(z_j=i|y_j,\theta_i^{(t)})
\end{equation}

In the above iteration relations Eq.\ref{gmm1}, Eq.\ref{gmm2} and
Eq.\ref{gmm3}, $t$ and $t+1$ denote the round of iterations. When we
ignore the measurement errors $\delta_j$, the above recursive
relation reduces to the standard EM recursive relation for Gaussian
Mixture Model. The above relations could be easily generalized to
multiviate case by simply substituting the data with data matrix,
mean with mean vector and variance with covariance matrix, which we
will not repeat the formula here. A C++ class that implements the
above algorithm is available upon request. }

\bibliography{maxridgebib}

\label{lastpage}

\end{document}